\documentclass[doublecol]{epl2} 

\title{Energy Fluxes during Dynamo Reversals}
\shorttitle{Title} 

\author{Pankaj Mishra\inst{1} \and Christophe Gissinger\inst{1} \and
  Emmanuel Dormy\inst{2} \and Stephan Fauve\inst{1}}
\shortauthor{P. Mishra \etal}

\institute{ \inst{1} Laboratoire de Physique Statistique, Ecole
  Normale Sup\'erieure, CNRS, Universit\'e P. et M. Curie,
  Universit\'e Paris Diderot, Paris, France\\ \inst{2}
  MAG(CNRS/ENS/IPGP), LRA, Ecole Normale Sup\'erieure, Paris,
  France\\ } 

\abstract{ Using direct numerical simulations of the equations of magnetohydrodynamics, we study reversals of the magnetic
field generated by the flow of an electrically conducting fluid in a sphere.  We show that at low magnetic Prandtl
numbers, $Pm=0.5$, the decrease of magnetic energy, ohmic dissipation and power of the Lorentz force during a reversal is
followed by an increase of the power injected by the force driving the flow and an increase of viscous dissipation. Cross correlations
show that the Lorentz energy flux is in advance with respect to the other energy fluxes. We also observe that
during a reversal, the maximum of the magnetic energy density migrates from one hemisphere to the other and comes
back to its initial position, in agreement with recent experimental observations. For larger magnetic Prandtl numbers ($Pm=$ 1, 2), the
magnetic field reversals do not display these trends and strongly differ one from another.}

\begin{document}

\maketitle

\section{Introduction}

It has been known since the work of Brunhes that the dipolar component of the magnetic field of the Earth has changed sign in a random way on
geologic time scale \cite{brunhes1906}. It is now believed that the Earth magnetic field is generated by a dynamo process, i.e. an instability
related to electromagnetic induction by the flow of liquid iron in Earth's core \cite{roberts2000}.  In the past 50 years, many models of
reversals of the magnetic field have been elaborated using various concepts and methods of fluid dynamics and dynamical system theory
\cite{petrelis2010}.  Although reversals are occurring randomly, a strong emphasis has been put on identifying patterns in the flow that
generate them and thus could be considered as precursors.  Parker first proposed two possible scenarios: the first one was related to
the fluctuations of the number and positions of cyclonic convective cells in the core \cite{parker1969}.  The second
one, based on a temporary weakening of the meridional circulation \cite{parker1979}, was indeed observed later in numerical simulations
\cite{sarson1999}.

Direct simulations of the magnetohydrodynamics (MHD) equations in a rotating sphere that display reversals of magnetic field have been achieved
since 1995 \cite{glatzmaier1995}.  Besides Parker's mechanism, other flow patterns have been identified as possible precursors of reversals,
such as convective plumes locally producing a magnetic field with opposite polarity \cite{wicht2004}. In more recent numerical simulations of the geodynamo, 
a stronger emphasis has been put on typical patterns of the magnetic field rather than the velocity field as possible precursors of field reversals \cite{aubert2008olson2011}. 
Note however that these simulations have been conducted at high values of $Pm$ (mostly in the range 10 to 20) and that it has been shown that the geometry 
of the magnetic field and its reversals strongly depend on the parameter values of these geodynamo numerical models \cite{busse2006kutzner2002}. 
On the one hand, no simple general pattern seemed to emerge since successive reversals in a given simulation can look different one from the other, 
but on the other hand, several simulations showed that the flow symmetries are playing an important role. The emission of poleward
light plumes identified as a precursor of reversals \cite{sarson1999,wicht2004}, breaks the equatorial symmetry of the flow in the liquid core.  
Breaking north-south symmetry of the convection pattern has been indeed found to be a necessary condition
for reversals in some geodynamo numerical models \cite{li2002nishikawa2008gissinger2012}.

This last feature has also been observed in a laboratory experiment involving a von Karman swiriling flow of liquid sodium driven by two
coaxial propellers in a cylinder (the VKS experiment).  With propellers counter-rotating at the same speed, only stationary dynamos were 
generated whereas counter-rotation at different speeds yields time-dependent regimes including random
reversals \cite{berhanu2007}. It has been shown that the symmetry broken by rotating the propellers at different speeds allows a linear
coupling between dipolar and quadrupolar modes of the magnetic field that provides a model for field reversals \cite{model}. Direct
simulations of the MHD equations with a similar flow forcing in a spherical geometry display the same features \cite{gissinger2010}.

Broken symmetries of the flow, either occurring spontaneously as in the geodynamo, or externally induced as in the VKS experiment, thus
play an important role in the observation of regimes involving reversals of the magnetic field.  The aim of this work is to identify
characteristic patterns of reversals with the help of numerical simulations. Instead of looking at patterns of the velocity or magnetic fields, 
we look at the behavior of energy transfers within the system during the process of field reversal. After recalling the governing equations, we will show how
the energy flux between the velocity and magnetic field, i.e. the power of the Lorentz force, is correlated with ohmic and viscous
dissipation as well as with the injected power by the forces driving the flow. We will then study the behavior of these energy fluxes
during reversals of the magnetic field.

\section {Governing equations and simulation details}

The MHD equations are numerically integrated in a spherical geometry
for the solenoidal magnetic ${\bf b}(r,\theta,\phi)$ and velocity
${\bf u}(r,\theta,\phi)$ fields
\begin{eqnarray}
 \frac{\partial \mathbf{u}}{\partial t}+Rm_0(\mathbf{u\cdot\nabla})\mathbf{u}&=& -Rm_0\mathbf{\nabla}\pi+Pm\triangle\mathbf{u} \label{eq:NDv}\\
             &&+Rm_0\mathbf{f}+ Rm_0\left({\bf b.\nabla}\right){\bf
   b}\, , \nonumber\\
\frac{\partial \mathbf{b}}{\partial t}&=&  Rm_0
\mathbf{\nabla\times}(\mathbf{u\times b}) + \triangle\mathbf{b}\, .\label{eq:NDb}
\end{eqnarray}

The above equations have been made dimensionless by using the radius
of the sphere $a$ as length scale and the magnetic diffusion time,
$\tau_0=\mu_0\sigma a^2$ as time scale. $\mu_0$ is the magnetic
permeability and $\sigma$ is the electrical conductivity.  $Pm=\mu_0
\sigma \nu$ is the magnetic Prandtl number where $\nu$ is kinematic
viscosity. $\pi$ is the pressure field.  The applied force is
$\mathbf{f}=f_0\mathbf{F}(r,\phi,z)$ where, $F_\phi =s^2\sin(\pi s
b)$, $F_z=\epsilon \sin(\pi sc)$ for $z>0$ and equal magnitude but
opposite sign for $z<0$. Polar coordinates ($s$,$\phi$,$z$) normalized
by $a$, are used for the representation of the forcing term. In order
to simulate finite sized impellers, this forcing is restricted to the
region $0.25 a < |z|< 0.65 a$ and $s<s_0$. Here, $s_0=0.4$, $b^{-1}=2s_0$,
and $c^{-1}=s_0$. This forcing term and non-dimensional form have both previously been used to model
both the Madison \cite{bayliss2007} and the VKS experiment \cite{gissinger2008,gissinger2010}. It is invariant by a rotation of
an angle $\pi$ along any axis in the midplane (herafter called the
$R_\pi$ symmetry). In order to reproduce the magnetic field reversals
observed in the VKS experiment, which only occur when the
counter-rotating impellers have different rotation rates, the $R_\pi$
symmetry in our simulations is broken by considering a forcing of the
form $C\mathbf{f}$, where $C$ is an asymmetry parameter fixed to $C=1$
for $z<0$ but can be different from one for $z>0$. A typical velocity
$V_0$ is used to define the input parameter $Rm_0=\mu_0\sigma
aV_0$. The magnetic Reynolds number is $Rm= max(u) Rm_0$ and the
kinetic Reynolds number is $Re=Rm/Pm$.

The above system of equations is solved using the Parody numerical code \cite{dormy1998}, originally developed for the
geodynamo and modified to make it suitable for configurations that involve a mechanical forcing of the flow.

We have performed simulations for different values of $Pm$ and $C$.  Depending on the value of these parameters, different dynamical regimes can be observed. 
For instance, when $Pm=0.5$, a transition from a statistically stationary axial dipolar magnetic field to chaotic reversals is observed for $C\ge 1.5$, in agreement
 with the numerical observations reported in ~\cite{gissinger2010}.

\section{Energy budget}

Eqs.~\ref{eq:NDv} - \ref{eq:NDb} are used to obtain equations for the
magnetic energy and the kinetic energy:
\begin{eqnarray}
 \frac{dE^{u}}{dt}&=& P-L-D^{u}, \label{eq:ke}\\ \frac{dE^{b}}{dt}&=&
 L - D^{b},\label{eq:me}
\end{eqnarray}
where, $ E^u = 1/2 \langle|\mathbf{u}|^2\rangle_V$ and $E^b =
1/2\langle |\mathbf{b}|^2\rangle_{\infty}$ are respectively the total kinetic
and magnetic energy, $P=
Rm_0\langle\mathbf{u\cdot}\mathbf{f}\rangle_V$ is the injected power,
$L = -Rm_0\langle\mathbf{u\cdot}[(\mathbf{\nabla\times b)\times
    b}]\rangle_V$ is the Lorentz flux, $D^u =
-Pm\langle\mathbf{u\cdot}\triangle\mathbf{u}\rangle_V$ is the viscous
dissipation, and $D^b =
-\langle\mathbf{b\cdot}\triangle\mathbf{b}\rangle_V$ is the ohmic
dissipation. In the above $\langle \cdot \rangle_V$ denotes spatial average over the sphere.

\begin{table*}
\begin{center}

\addtolength{\tabcolsep}{-3pt}
\caption{Space and time averaged values of injected power ($P$),
  viscous ($D^{u}$) and ohmic ($D^{b}$) dissipations, Kinetic ($E^u$)
  and magnetic ($E^b$) energies for different values of $Pm$ and the
  asymmetry parameter $C$. For $Pm=0.5$, the injected power increases
  with $C$, whereas the ratio of kinetic to magnetic energy
  decreases.} 
\vspace{5mm}
\begin{tabular}{|c|c|c|c|c|c|c|c|c|c|c|c|c|}

\hline 
$Pm $& $C$ & $Rm_0$& $Re$ & $P$ & $D^{u}$ & $D^{b}$ & $L $& $E^u$ & $E^b$ & $E^u/E^b$ & $D^{u}/P$ & $D^{b}/P$\tabularnewline
\hline 
\hline 
 & 1 &300 & 230 & 3.99 & 3.44 & 0.59 & 0.588 & 0.0112 & 0.00067 & 16.7 & 0.86 & 0.14\tabularnewline
\cline{2-13} 
 & 1.05 &300 & 240 & 4.09 & 3.48 & 0.62 & 0.62 & 0.0112 & 0.00070 & 16.0 & 0.85 & 0.15\tabularnewline
\cline{2-13} 
 & 1.25 &300 & 237 & 4.46 & 3.68 & 0.78 & 0.78 & 0.0114 & 0.00085 & 13.4 & 0.82 & 0.18\tabularnewline
\cline{2-13} 
0.5 & 1.5 &330& 280 & 5.60 & 4.66 & 1.04 & 1.03 & 0.0141 & 0.0011 & 12.8 & 0.83 & 0.17\tabularnewline
\cline{2-13} 
 & 2.0 & 390 &370 & 8.17 & 6.24 & 2.06 & 2.05 & 0.0166 & 0.0079 & 2.1 & 0.76 & 0.25\tabularnewline
\hline 
1.0 & 2.0 &700& 400 & 15.32 & 10.86 & 4.78 & 4.73 & 0.0148 & 0.0034 & 4.4 & 0.70 & 0.30\tabularnewline
\hline 
2.0 & 2.0 &1000& 320 & 21.11 & 12.61 & 8.60 & 8.59 & 0.0088 & 0.0062 & 1.4 & 0.60 & 0.40\tabularnewline
\hline 
\end{tabular}
\end{center}

\label{table:eb}
\end{table*}

Eq.~\ref{eq:ke} expresses that the rate of change of the kinetic energy is equal to the difference between the power injected by the
driving force and the combined kinetic energy loss due to both the viscous dissipation
and the power of the Lorentz force (hereafter called the Lorentz flux). $L$ is here defined such that $L>0$ corresponds to a
positive transfer of energy from the velocity field to the magnetic field. Eq.~\ref{eq:me} shows that the rate of change of the magnetic
energy is the difference between the Lorentz flux and ohmic dissipation. Therefore, the Lorentz flux acts as a source term for the
magnetic energy.

The mean values of the energies and energy fluxes that have been computed for different values of the governing parameters are reported
in table 1.

Several observations can be made from the Table 1. First, note that in all our runs, the ratio between
the kinetic and the magnetic energy is always greater than one, and seems to be controlled by the magnetic Reynolds number, or more
exactly by the distance from dynamo onset $Rm-Rm_c$, where $Rm_c$ is the dynamo onset. Indeed, for a fixed value $Pm=0.5$, this ratio
increases as $Rm$ is decreased, $E^{u}$ and $E^b$ being of the same order of magnitude when $Rm$ is much larger than $Rm_c$.  In most 
of the runs, the main part of the injected power is dissipated by viscosity, as the ratio $D^u/D^b$ is generally greater than $0.5$. $D^b/D^u$ 
increases as $Rm$ or $Pm$ are increased.

Global energy balances have already been used to analyze geodynamo models \cite{buffet2002} but mostly in statistically stationary
regimes. The emphasis will be placed here on fluctuations of the energy fluxes and their correlations, in particular during field reversals.
 
\begin{figure}[htbp]
\onefigure[trim=7mm 0mm 1.6cm 1.2cm, scale=0.33]{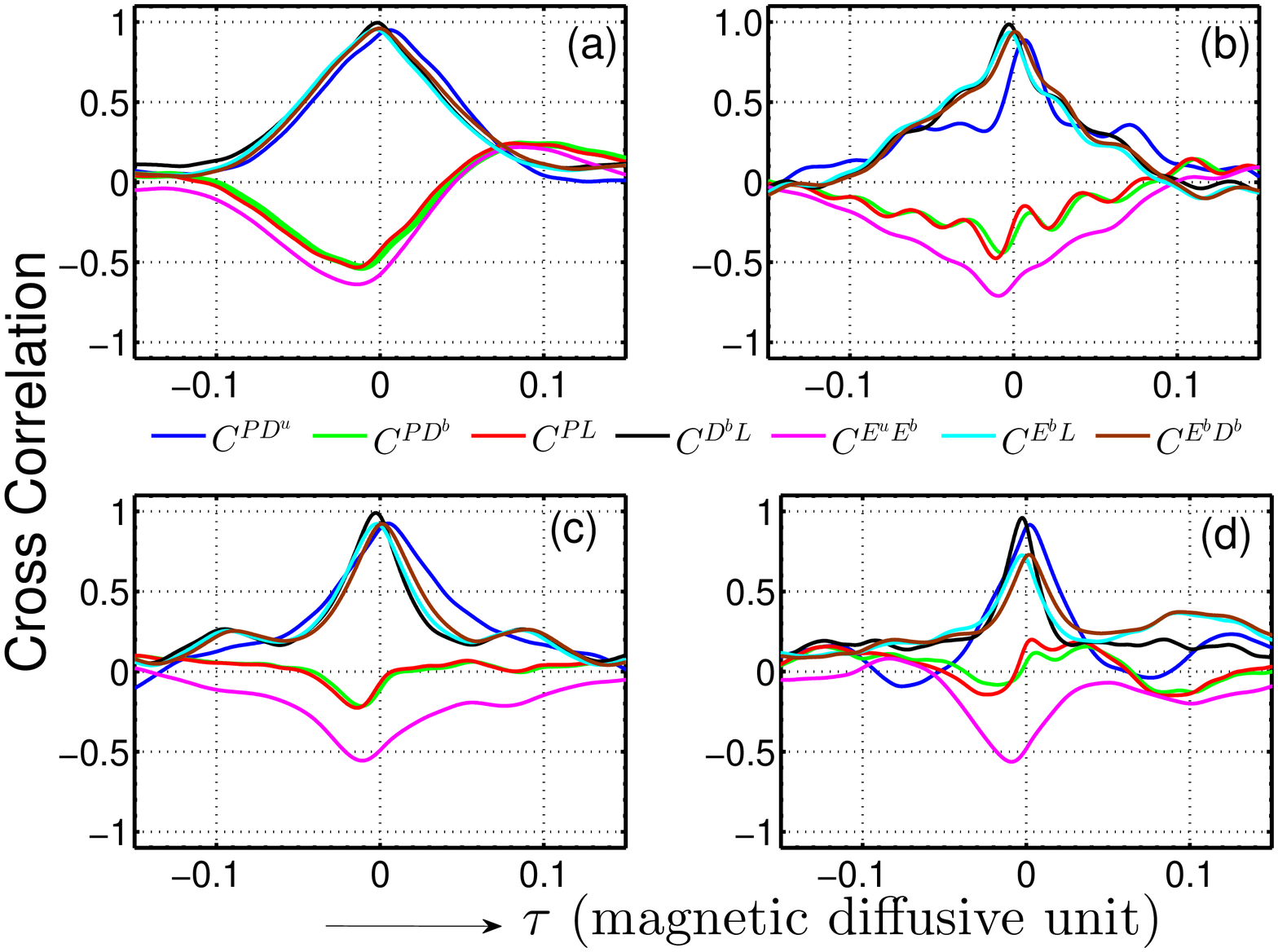}
\vskip -5 mm
\caption{Plot of the cross correlation $C^{XY}(\tau)$ versus delay
  time $\tau$ for : (a) $Pm=0.5$, $C=1.5$ ($Rm_0=330$), (b) $Pm =
  0.5$, $C=2$ ($Rm_0=370$), (c) $Pm = 1$, $C=2$ ($Rm_0=700$), and (d)
  $Pm = 2$, $C=2$ ($Rm_0=1000$). }
\label{fig:cross_cor}
\end{figure}

Although the flow is driven by a constant force, all the spatially averaged quantities involved in the energy budget
(Eqs.~\ref{eq:ke}-\ref{eq:me}) fluctuate in time since both the velocity and magnetic fields are chaotic.  We consider the cross
correlation functions between these quantities in order to get some insight on their fluctuations.  The cross correlation between two
variables $X$ and $Y$ is defined as $C^{XY}(\tau)~=~\langle (X(t)-\bar{X})(Y(t+\tau)-\bar{Y})/{\sigma_X}^2 {\sigma_Y}^2\rangle$,
where ${\sigma_X}^2~=~(\overline{X^2(t)}-\bar{X}^2)$, and $\sigma_Y~=~(\overline {Y^2(t)}-\bar{Y}^2)$ are the variance of $X$
and $Y$ respectively. $\bar{X}$ and $\bar{Y}$ represent mean values.  The cross correlation functions between
all the terms in Eqs.~\ref{eq:ke}-\ref{eq:me} computed from our numerical simulations are shown in Fig.~\ref{fig:cross_cor}. 

We first observe that purely kinetic quantities, $P, D^{u}$ and $E^{u}$ (not displayed), are well correlated, the maximum of the cross
correlation being close to $1$. Similarly, the Lorentz flux is well correlated with both the magnetic energy $E^{b}$ and ohmic dissipation
$D^{b}$ although it is not a purely magnetic quantity (it involves the velocity field). A weaker correlation is observed
between the purely kinetic quantities and quantities involving the magnetic field: the amplitude of their cross correlation peaks at
about $0.5$. In addition, the correlation between the injected power and the Lorentz flux (or ohmic dissipation) decreases when the
magnetic Prandtl number increases (see Fig.~\ref{fig:cross_cor} (c), (d)). The source terms for kinetic (resp. magnetic) energy are thus
less correlated when $Pm$ increases.

Another feature is related to the sign of the extremum of the cross correlation function. It is positive among the quantities involved in
the magnetic energy budget (Eq.~\ref{eq:me}), such that a fluctuation in the Lorentz flux is on average followed by a
fluctuation with the same sign for both magnetic energy and ohmic dissipation.  In contrast, it is followed by a fluctuation of the
opposite sign for the injected power. Similarly, the kinetic and magnetic energies are anti-correlated, a decrease in magnetic energy
being followed on average by an increase of kinetic energy.

One last information is provided by the time lags displayed by the correlation functions. It can be observed from
$C^{PD^{u}}(\tau)$ that there is a time lag between injected power and viscous dissipation (peak at positive delay time). This corresponds to
the time needed for the kinetic energy to cascade to dissipative scales, as often observed in turbulent flows
\cite{lumley1992}.  Similarly, the Lorentz flux that is the source term for the magnetic energy, is slightly in advance compared to the
magnetic energy and ohmic dissipation.  Less intuitive are the cross correlations that involve both kinetic and magnetic quantities. It is
indeed observed that the Lorentz flux, magnetic energy and ohmic dissipation are all in advance compared to the injected power and the
other kinetic quantities. The above results therefore suggest that the fluctuations of the Lorentz flux are the ones that
coherently affect the other energy fluxes after some time lag. This can be understood since the Lorentz flux couples kinetic and
magnetic modes, thus its fluctuations are likely to affect the balance between kinetic and magnetic energy in a coherent way. Although
it is clear from Eqs.~\ref{eq:ke}-\ref{eq:me} that a fluctuation of $L$ will affect $E^{u}$ and $E^{b}$ in opposite ways, we however
emphasize that it is not intuitive that the fluctuations of the kinetic energy lag the ones of the magnetic energy. This may depend on
the value of $Pm$ (out of the range of the present study) or on the way the flow is forced.

The above results strongly suggest a precise chronology in the energy fluxes during the magnetic field dynamics. In the following section,
we therefore discuss their behavior during magnetic field reversals.

\section{Energy fluxes during dynamo reversals}

\begin{figure}[htbp]
 \begin{center}
 \onefigure[trim=15mm 3mm 1.6cm 2mm, scale=0.34]{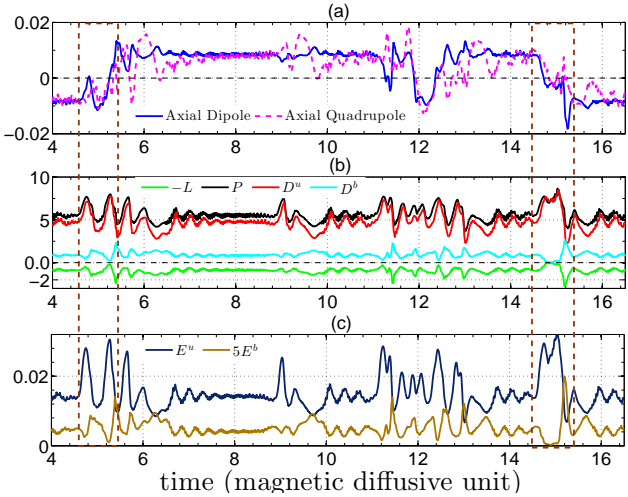}
 \end{center}
\vskip -5 mm
\caption{Energy fluxes for $Pm=0.5$, and $C=1.5$.  Top panel: time series of axial dipole (blue) and quadrupole (pink).
Middle panel: Lorentz flux $L$ (green), injected power $P$ (black), viscous dissipation $D^u$ (red) and ohmic dissipation
$D^b$ (cyan).  Bottom panel: kinetic $E^u$ and magnetic $E^b$ energies.}
\label{fig:eb_Pmp5_C1p5}
\end{figure}

Fig.~\ref{fig:eb_Pmp5_C1p5} shows the temporal dynamics of the
injected power $P$, dissipation rates ($D^u$ and $D^b$), and Lorentz
flux $L$ for $Pm=0.5$, and $C=1.5$, when chaotic magnetic field
reversals are observed. At the beginning of a reversal, as the
amplitude of the axial dipole decreases, both the Lorentz flux and
ohmic dissipation decrease. Since ohmic dissipation stays larger than
the Lorentz flux, this phase is associated with a weakening of the
magnetic energy.

On the contrary, in the kinetic equation, the injected power $P$ and the viscous dissipation $D^u$ both increase when the dipole
vanishes. Since the net dissipation $D^u+L$ in the kinetic equation stays small compared to the total injected power, the kinetic energy
thus increases during this period.

Once the dipole starts recovering its mean value, both the Lorentz flux and ohmic dissipation increase such that the magnetic energy
grows.  Viscous dissipation and injected power decrease in producing a net kinetic dissipation, i.e. $L+D^u$ is larger than the total
injected power such that there is a decrease of the total kinetic energy during the recovery of the dipolar field. Once the reversal is
over, all the quantities start fluctuating again along their mean value.

\begin{figure}[htbp]
 \begin{center}
 \onefigure[trim=15mm 3mm 1.6cm 2mm, scale=0.34]{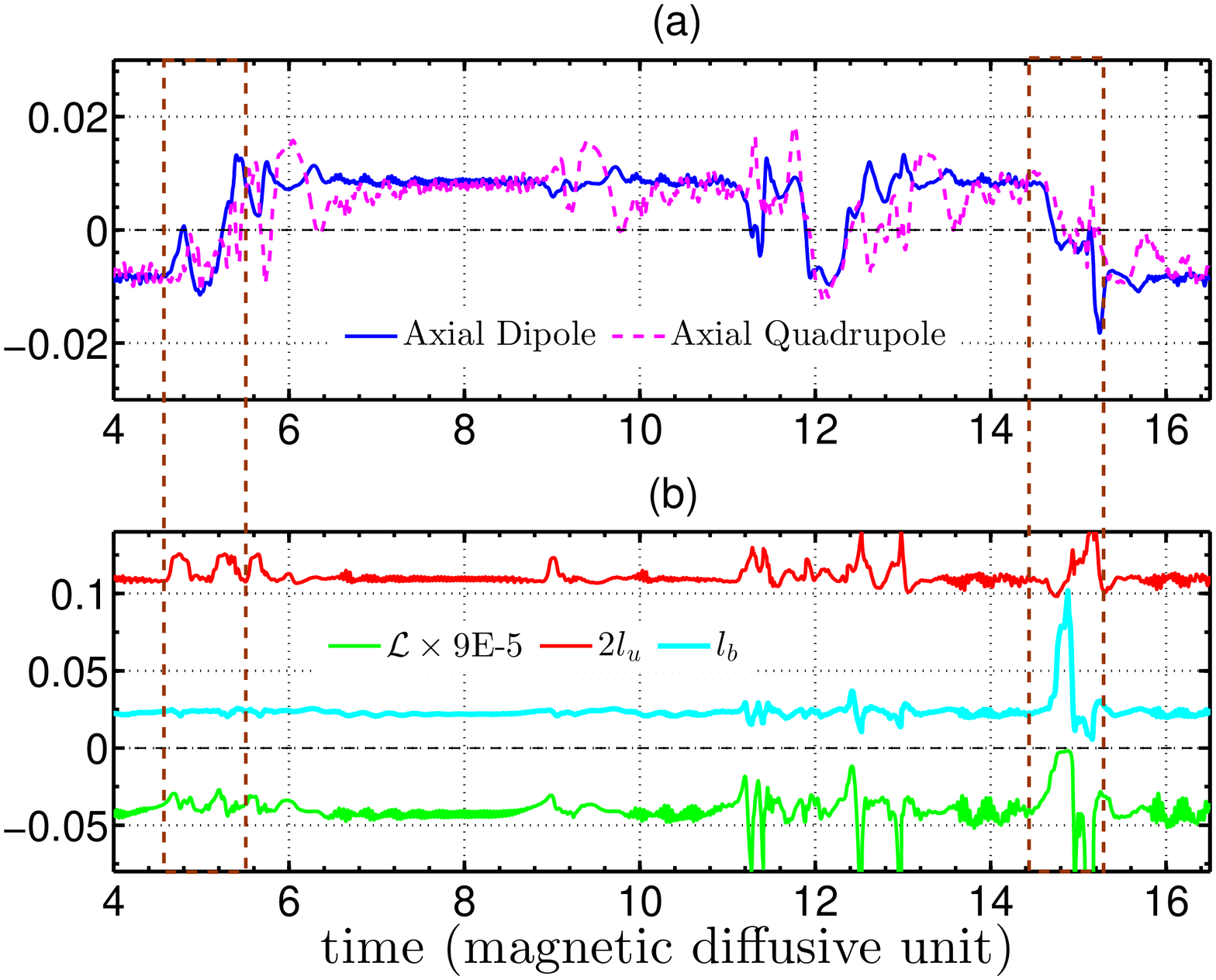}
\end{center}
\vskip -5 mm
 \caption{Top panel: time series of axial dipole (blue) and axial quadrupole (pink).  Bottom panel: time series of the normalized
   Lorentz flux $\mathcal{L}$ (green), effective velocity length scale
   $l_u$ (red) and effective magnetic length scale $l_b$ (cyan), for
   $Pm=0.5$, and $C=1.5$.
 }
 \label{fig:dyn_length}
 \end{figure}

Thus, for the simulations made at $Pm=0.5$, the properties of the energy fluxes inferred from the cross correlation functions computed
on the whole time recordings can be observed on each individual reversal. The power of the Lorentz force decreases at
the beginning of a reversal followed by the decrease of the magnetic energy or ohmic dissipation, whereas the injected power, the kinetic
energy and then viscous dissipation first increase.  In contrast to the kinetic and magnetic energies, the energy fluxes involve spatial
derivatives of the velocity or magnetic fields. Thus, the fluctuations of viscous (resp. ohmic) dissipation can be related to fluctuations of
the amplitude of the velocity (resp. magnetic) field and to change in their characteristic length scale.  Consequently, it is not clear
whether the decrease (resp. increase) of ohmic (resp. viscous) dissipation is primarily related to the decrease (resp. increase) of
the magnetic (resp. kinetic) energy. This is also true for the Lorentz flux, $L = -Rm_0\langle\mathbf{u\cdot}[(\mathbf{\nabla\times b)\times b}]\rangle_V$, 
that can be also written $L = -Rm_0\langle (\partial_i u_j + \partial_j u_i) b_i b_j \rangle_V \,/2$ if the flow
is incompressible. An additional feature of the Lorentz flux is its dependence on the respective angles between the different fields.  In
order to get further insight in the fluctuations of the energy fluxes during reversals, we normalize the viscous (resp. ohmic) dissipation
by the kinetic (resp. magnetic) energy, by computing the quantities $l_u=[D^u/E^u)]^{-1/2}$ and $l_b=[D^b/E^b]^{-1/2}$.  In addition we
also define the normalized Lorentz flux as $\mathcal{L}= L/\sqrt{D^u}E^b$. Note that these normalized dissipations can be
regarded as changes in the kinetic (resp.magnetic) length scale, while the renormalized Lorentz flux involves some correlation between local
dissipation and the magnetic field.  Fig.\ref{fig:dyn_length} displays these quantities. The velocity length scale increases when reversals occur 
and strongly decreases during the dipole overshoot. The behavior of the magnetic length scale and of the normalized Lorentz flux 
displays more variability from one reversal to the other. This results from the variability of the minimum magnetic energy achieved during reversals 
(see Fig.~\ref{fig:eb_Pmp5_C1p5} (c)).

\begin{figure}[htbp]
 \begin{center}
\onefigure[trim=15mm 3mm 1.6cm 2mm, scale=0.34]{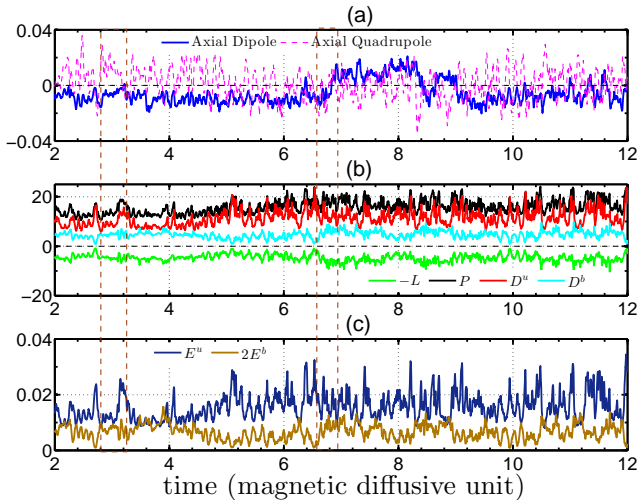}  
\end{center}
\vskip -5 mm
\caption{Energy fluxes for $Pm=1.0$, and $C=2.0$.  Top panel: Time series of axial dipole (blue) and quadrupole (pink).  Middle panel: 
Lorentz flux $L$ (green), injected power $P$ (black), viscous dissipation $D^u$ (red) and ohmic dissipation $D^b$ (cyan).  Bottom
panel: kinetic energy $E^u$ and magnetic energy $E^b$. Compared to $Pm=0.5$, most of the energy
fluxes show no systematic behaviour. For some reversals, (e.g., $t\sim7$) there is an increase of $L$ and $D^b$ as dipole vanishes,
which contrasts with the dynamics observed for Pm=0.5.}
 \label{fig:eb_Pm1_C2}
 \end{figure}

The characteristic features displayed by the energy fluxes during field reversals for $Pm=0.5$ are not observed for $Pm$ larger ($Pm=1$
or $2$), as shown by the different time recordings in Fig.~\ref{fig:eb_Pm1_C2}. Neither the kinetic (resp. magnetic)
energy nor the different energy fluxes display some systematic changes during reversals compared to regimes of given polarity. The effect of
$Pm$ on the reversal mechanism has been emphasized in previous studies \cite{gissinger2010}: it has been shown that the reversals obtained at
$Pm=0.5$ primarily involve the coupling of the axial dipole with an axial quadrupole in the framework of a simple model \cite{model}
whereas more modes generate dynamics with more variability for reversals observed at larger $Pm$.

\section{Drift of the magnetic energy during field reversals}

\begin{figure}[htbp]
 \begin{center}
 \onefigure[trim=15mm 3mm 1.6cm 2mm, scale=0.34]{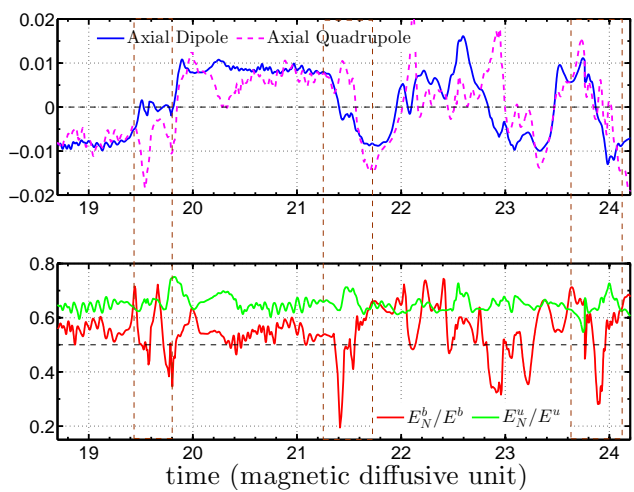}
\end{center}
\vskip -5 mm 
\caption{Top Panel: axial dipole (blue) and axial quadrupole (magenta)
  for $Pm=0.5$ and $C=1.5$. Bottom panel: Time series of the relative
  kinetic energy in the northern hemisphere $E^u_N/E^u$ (green) and of
  the relative magnetic energy in the northern hemisphere $E^b_N/E^b$
  (red).}
 \label{fig:eb_local_Pmp5_C1p5}
 \end{figure}

It has been recently reported in the VKS experiment that a spatial
localization of the dynamo magnetic field occurs when the symmetry of the driving is broken by rotating the two propellers
at slightly different velocities \cite{gallet2012}. In addition, it has been observed that the maximum of magnetic energy
drifts during a field reversal. It crosses the equatorial plane and comes back to its initial location at the end of the reversal. These
phenomena could be related to hemispherical dynamos observed in some planets or stars. The present numerical simulations provide
a convenient tool to study hemispherical dynamos and the spatiotemporal dynamics of the magnetic field during reversals.  To wit, we compute the kinetic 
(resp. magnetic) energy in the northern hemisphere $E^{u}_N$ (resp. $E^{b}_N$) by integrating the kinetic (resp. magnetic) energy density restricted to the 
northern hemisphere. The kinetic (resp. magnetic) energy fraction in the northern hemisphere $E^u_N/E^u$ (resp. $E^b_N/E^b$) during field reversals is 
displayed in Fig.~\ref{fig:eb_local_Pmp5_C1p5}.  We observe that $E^u_N/E^u$ is significantly larger than $0.5$ because the flow forcing is stronger in the 
northern hemisphere ($C=1.5$). The magnetic field is also on average stronger in the northern hemisphere. However, $E^b_N/E^b$ drops significantly below $0.5$
during field reversals. Thus, the maximum of magnetic energy density migrates from the northern to the southern hemisphere
during a field reversal and then comes back to its  initial location after the reversal.

Some differences with the VKS experiment should be mentioned: in the experiment, the magnetic energy is stronger
close to the slow propeller when the speeds are slightly different. It becomes localized close to the fast propeller, similarly to the present
simulations, only when the rotation speeds significantly differ. In addition, the asymmetry in the distribution of the magnetic energy
density in the experiment looks stronger than in the simulations. This can be related to the difference of thresholds of the dipolar and quadrupolar modes
\cite{gallet2009,gallet2012}. Reversals in the VKS experiment can be
superimposed whereas they display a much stronger variability in the
present numerical simulations, even at the lowest value of $Pm$. This
is related to the stronger contribution of modes higher than the
dipolar and quadrupolar ones in the simulation. Despite these higher
modes, the diagnostic provided by the recording of $E^b_N/E^b$ during
reversals is rather robust when $Pm$ is small enough and the migration
of the maximum magnetic energy density through the equatorial plane
during reversals is in agreement with the experimental observations.


\section{Conclusion} 

We have studied the properties of energy fluxes and their correlations in fluctuating dynamo regimes involving field reversals.  Cross
correlation functions show that fluctuations of the power of the Lorentz force are in advance with respect to fluctuations of ohmic and
viscous dissipation as well as magnetic or kinetic energy. Somewhat surprisingly, kinetic quantities lag behind the magnetic ones. The
magnetic and kinetic energy fluctuations are anti-correlated. It will be worth checking the robustness of these findings in other
dynamos. In particular, the role of the Lorentz flux as a possible precursor of dynamo fluctuations deserves further
studies. To the best of our knowledge, geometrical properties such as the distribution of the angle between the velocity and the magnetic
(resp. current density) field, and their contribution to the fluctuations of the Lorentz flux, have not been investigated in
turbulent dynamos.

The study of fluctuations of the energy fluxes confirmed that simple patterns during field reversals are observed only when the magnetic
Prandtl number is small enough. This also deserves to be checked on other dynamo numerical models.  We observe that for $Pm =0.5$, as the
reversal starts, the Lorentz flux and and then magnetic energy and ohmic dissipation decrease, while injected power, kinetic energy and
then viscous dissipation increase. The magnetic energy decreases while the kinetic energy increases until the dipole vanishes.  Once the
dipole starts recovering and overshoots to the other polarity, the opposite behavior is observed.

Another diagnostic used in this study for accessing spatiotemporal dynamics of field reversals, consists in recording the fraction of
magnetic energy in one hemisphere.  When reversals mainly involve an interaction between modes with dipolar (resp. quadrupolar) symmetry,
this recording shows that the maximum magnetic energy density crosses the equatorial plane and then comes back to its initial position during
each reversal.

The above features are not observed for reversals occurring when $Pm$ is larger ($Pm =1$ and $2$) that display more variability due to the
interplay of more magnetic modes.  Although the range of $Pm$ that can be scanned in direct numerical simulations is very far from realistic
values for liquid metals, this trend is consistent with the very reproducible field reversals reported in the VKS experiment for which
$Pm \simeq 10^{-5}$.

\acknowledgments
Support of IFCPAR/CEFIPRA contract 4904-A is acknowledged.

\end{document}